\documentclass[sigconf]{acmart}
\AtBeginDocument{%
  \providecommand\BibTeX{{%
    \normalfont B\kern-0.5em{\scshape i\kern-0.25em b}\kern-0.8em\TeX}}}

\setcopyright{none}

\begin{document}

\title{A Survey of Data Security: Practices from Cybersecurity and Challenges of Machine Learning}


 \author{Padmaksha Roy}
  \affiliation{%
   \institution{Virginia Tech}
   \state{Virginia}
   \country{USA}
 }
  \author{Jaganmohan Chandrasekaran}
  \affiliation{%
   \institution{National Security Institute \\ Virginia Tech}
   \state{Virginia}
   \country{USA}
 }
 \author{Erin Lanus}
\authornote{Corresponding author lanus@vt.edu}
 \affiliation{%
   \institution{National Security Institute \\ Virginia Tech}
   \state{Virginia}
   \country{USA}
 }
 \author{Laura Freeman}
 \affiliation{%
   \institution{National Security Institute \\ Virginia Tech}
   \state{Virginia}
   \country{USA}
 }
  \author{Jeremy Werner}
 \affiliation{%
   \institution{Director Operational Test and Evaluation (DOT\&E)}
      \country{USA}
 }

\renewcommand{\shortauthors}{Roy, Chandrasekaran, Lanus, Freeman, and Werner}

\begin{abstract}
Machine learning (ML) is increasingly being deployed in critical systems. The data dependence of ML makes securing data used to train and test ML-enabled systems of utmost importance. While
the field of cybersecurity has well-established practices for securing information, ML-enabled systems create new attack vectors. Furthermore, data science and cybersecurity domains adhere to their own set of skills and terminologies. This survey aims to present background information for experts in both domains in topics such as cryptography, access control, zero trust architectures, homomorphic encryption, differential privacy for machine learning, and federated learning to establish shared foundations and promote advancements in data security.

\end{abstract}



\maketitle
{\verb||}

\section{Introduction}
With the abundance of data in the digital age, organizations are finding new ways to create, store and transfer data at an ever-growing pace which has called for best practices in data governance to be formalized. IBM defines data security as ``the practice of protecting digital information from unauthorized access, corruption, or theft throughout its entire life cycle'' \cite{datasec}. Data is an essential component of information systems, and data security issues need to be considered at each level of the system. That is, data resides on physical devices, is manipulated by software processes, and is foundational for most applications in autonomy and artificial intelligence (AI), such as public health and autonomous vehicles. Thus, data security is informed by many aspects of cybersecurity, ranging from physical security of data storage, network security over which data travels, cryptographic algorithms to encrypt data and cryptographic protocols to achieve authentication, logical aspects of access control mechanisms, to social, legal, and administrative policies for data use and governance. 

A subfield of AI, machine learning (ML) is particularly data dependent and is increasingly included  as a component of information systems with many opportunities along the ML lifecycle -- ranging across data collection, preprocessing, training, testing, and deployment -- for data to be adversarially or accidentally compromised. In addition to incorporating established cybersecurity best practices for protecting information in data management platforms, new challenges to security and privacy presented by ML pipelines need to be addressed. The goal of this survey is to inform the data scientist and ML practitioner of existing cybersecurity mechanisms that are relevant to data security and to highlight security and privacy concerns presented by the deployment of ML for the cybersecurity expert in order to establish a common foundation for discussion between these two groups and promote advancements in data security.

\section{Cybersecurity Approaches to Data Security}
\subsection{Security and Privacy Concepts}
The cybersecurity of an information system is most commonly discussed in terms of three foundational security properties -- confidentiality, integrity, and availability -- denoted the ``CIA triad.'' 
In this context, confidentiality is defined as ``preserving authorized restrictions on information access and disclosure including means for protecting personal privacy and proprietary information'' \cite{NIST_CONFIDENTIALITY}, ensuring that ``sensitive information is not disclosed to unauthorized entities.'' Integrity guards against ``improper information modification or destruction and includes ensuring information non-repudiation and authenticity.'' It guarantees the data ``have not been altered in an unauthorized manner since it was created, transmitted, or stored'' \cite{NIST_INTERGITY}, and ensures that such changes to the data are detectable. Cryptographic techniques and access control are both utilized to ensure confidentiality and integrity of data, although in different ways. Availability means ensuring timely and reliable access to information, the specifics of which are dependent on how the information is being retrieved. Ensuring availability typically requires protections to communications networks to ensure transit of information and to end points to ensure servers have available resources to meet legitimate client requests, as well as redundancy of storage to fulfill requests when access to one set of the data is denied or the data itself is compromised. 

Confidentiality of data is most commonly concerned with the proprietary or sensitive data owned by an organization, such as a corporation or a government agency. Privacy, on the other hand, is the ``assurance that the confidentiality of, and access to, certain information about an entity is protected'' and ``the freedom from intrusion into the private life or affairs of an individual when that intrusion results from undue or illegal gathering and use of data about that individual'' \cite{NIST_PRIVACY}. While access control and cryptographic techniques can guarantee the secrecy of the private data against unauthorized exposure in the same way that they guarantee secrecy of confidential data, there is more to achieving privacy than secrecy. Typically, privacy is measured with respect to personally identifiable information which is defined as ``information that can be used to distinguish or trace an individual’s identity, either alone or when combined with other information that is linked or linkable to a specific individual'' \cite{SP800633}. Some aspects of privacy protection are legal or policy-based, specifying what information a corporation or organization can collect about an individual, with whom it can be shared, and the rights of individuals to control their information and to be informed about its use. Enforcing privacy rules in data typically requires a technical solution, and these have evolved over time along with our information systems. Past approaches have relied on restricting the kinds of queries that can be posed to a database containing PII or anonymizing the data prior to disclosure, but correctly identifying the private attributes that must be protected becomes crucial. Differential privacy is the current state-of-the-art, and it also has applications in protecting private training data on which machine learning-enabled systems operate.

Different systems require different levels of assurance and have different security goals and properties that must be upheld. Generally, the security properties determined for a given system are matched
to the sensitivity of the information needing to be protected, the budget of the defenders,
the expected compliance of the legitimate users, and the anticipated adversary model of the
attackers. In government systems, levels of assurance can be categorized based on the the level of impact to the confidentiality, integrity, and availability security properties. Impacts are classified as high, moderate and low depending on severity. The level determined is based on the high-water mark or worst case impact among the CIA triad security objectives for all information contained within the system. The minimum security requirements outlined in NIST FIPS publication 200 cover seventeen security-related areas with regard "to protecting
the confidentiality, integrity, and availability of federal information systems and the information processed, stored, and transmitted by those systems"  \cite{pub2005minimum}. In this work, we focus on the listed security-related areas: (i) access
control; (iii) audit and accountability; (vii) identification and authentication; (xv) systems and services acquisition; (xvi) system and communications protection; and (xvii) system and information integrity.

The NIST Cybersecurity Framework provides five core functions that can be employed concurrently and iteratively by organizations to assess their cybersecurity risk \cite{NIST_CSWP}. These core functions include identification of critical assets and risks, protecting assets through development of safeguards, detecting attacks via alerts and monitoring, responding to detected attacks to contain the impact, and recovering assets and services back to normal operation. The concept of defense-in-depth is that a single protection mechanism should not be responsible for the security of an information system as almost no protection mechanisms provide unconditional security against every attacker \cite{SP80053r5}. Each protection mechanism layered in a defense-in-depth strategy is put in place to increase the computational budget needed by the attacker, either making it computationally infeasible for an attacker to succeed as a form of deterrence or to slow down the attacker allowing the defenders time to detect and respond. Defense-in-depth strategies consider multiple dimensions of the organization including both people and technology. 

\subsection{Cryptography}
Encryption enables confidentiality, ensuring that the plaintext is only accessible by parties with access to decryption keys. To further ensure that only authorized parties can observe the plaintext requires the use of secure cryptographic protocols and keys that are correctly distributed and managed.

Secure cryptographic hash functions are one-way functions, $h(m)$, that should have three properties: preimage resistance, collision resistance, and second preimage resistance. A one-way function has preimage resistance; that is, a preimage $m$ cannot be computed from a randomly chosen image $h(m)$. Collision resistance means it is computationally infeasible to find two preimages, $m$ and $m'$ that hash to the same randomly chosen image, $h(m) = h(m')$. Second preimage resistance means it is computationally infeasible to compute a second preimage $m'$ given a specific preimage $m$ that hash to the same image, $h(m) = h(m')$. The computational work to compute a collision is exponential in the strength of resistance and is generally half the length of the hash, though it can also depend on the length of the message \cite{SP800107}. Broken hash algorithms are those for which attacks against some of the the security properties have been demonstrated. Currently, this list includes Message Digest 5 (MD5) and Secure Hash Algorithm 1 (SHA-1).

Encryption can be broadly divided into symmetric-key and asymmetric-key encryption.  In symmetric-key encryption, both sender and receiver share the same key $k$ and  one function encrypts the plaintext into the ciphertext $e(k,p) = c$ while another function performs the inversion to decrypt the ciphertext back into the plaintext, $d(k,c) = d(k,e(k,p)) = p$. In asymmetric-key encryption, there is only one encryption function, $e(k, m) = m'$ but two keys which are the inverses of each other, $k$ and $k^{-1}$. Encrypting a plaintext $p$  with one key, say $k$, creates a ciphertext $e(k,p) = c$ which is invertible by encrypting with the other key, $e(k^{-1},e(k,p)) = p$. 

Symmetric-key encryption is generally faster and less resource intensive than asymmetric-key encryption, so it often serves to provide confidentiality on large amounts of data or once a shared secret, such as a session key, has been established. It does require a secure channel for sharing the key or for the key to be pre-shared. Additionally, protocols or procedures using symmetric-key cryptography must differentiate encryption and decryption functions. Further, a message encrypted with the key could have been produced by any of the participants with the shared secret. 

The inverse key nature of asymmetric-key encryption makes it useful for other  services. When a single party maintains one of the keys as private and publishes the inverse key as public, this enables public key encryption and digital signatures. The public key is often signed by a trusted third party known as a certificate authority in systems with public key infrastructure (PKI), and this certificate serves to bind the public key to the entity that holds the private key. Anyone with the certificate can use the public key to encrypt a message for the private key holder, but this message can only be decrypted by the private key holder. Typically, the private key holder encrypts a hash of a message with the private key and sends it along with the message. This signed hash can be decrypted by anyone with the public key, but could only have been encrypted by the private key; thus, the signed hash is the private key holder's signature on the sent message \cite{DS_NIST}.

Cryptographic protocols combine primitives in different ways to achieve different goals. The practice of encrypting information with one key and encrypting that key with another key is known as envelope encryption. Symmetric keys are often used as the information-encrypting key due to the benefit of using the faster algorithm on the large amount of information while asymmetric keys may be used to encrypt the symmetric key due to the benefits provided by asymmetric cryptography. For example, encrypting a symmetric key with a public key can be used to start a communication session that authenticates the receiver, the private key holder, to the sender, but does not achieve mutual authentication. Diffie-Hellman key exchange is a public-key protocol using asymmetric encryption that can be used to compute a shared secret without authentication of either party. In either case, once the symmetric key is shared, confidential communication between the two parties is achieved via symmetric-key encryption where long term use of asymmetric encryption would be too slow and computationally burdensome.

\subsection{Authentication}  
The most basic and common definition of authentication is the process of ``verifying the identity of a user, process, or device, often as a prerequisite to allowing access to resources in an information system'' \cite{pub2005minimum}. Most commonly, authentication refers to user authentication, such as when a client authenticates to a server. The three common authentication factor categories used to prove identity are:
\begin{enumerate}
    \item something you know, such as a password;
    \item something you have, such as a credential card, cryptographic key fob,  or mobile phone authenticator app; and
    \item something you are, typically involving biometric factors such as a fingerprint, retina, or face \cite{NIST_SFA}.
    \end{enumerate}

A system utilizing single factor authentication is satisfied by proof of identity using only one factor in a session \cite{NIST_SFA}. Each factor has vulnerabilities that require mitigation strategies, which is particularly critical when a single factor is the only obstacle to an adversary seeking to impersonate the entity. Passwords can be one of the most vulnerable forms of user authentication based on how they are used, so we provide examples of password-based authentication vulnerabilities and the best practices that aid in mitigation. Early flawed protocols sent passwords in the clear, and storing passwords in plaintext on the server makes it possible for an attacker to steal passwords by gaining access to the server. Storing the hash of passwords using a secure hashing algorithm allows for verification of a submitted password against the stored hash, while the one-way hash function prevents the attacker from obtaining the password itself. Rainbow tables are a technique of precomputing password hashes for a dictionary of passwords to reduce the computational time of password cracking, but are mitigated by adding a salt, a unique random value when computing the hash for an individual's stored password hash, which increases the space of hashed values. Brute force attacks attempt to guess a password by systematically trying all possible combinations; increasing the length of character set through the use of numbers and symbols increases the search space. Locking an account after a number of failed attempts as a mitigation strategy can actually become a vector for a denial of service attack, but other strategies that increase the computational budget and thus slow down the attacker include CAPTCHAs to prevent automation and requiring complex passwords to increase the search space \cite{bruteforce}. 

The previous examples of attacks focus on cryptographic vulnerabilities and software mitigations, but humans are another source of vulnerabilities. Password reuse across multiple platforms increases vulnerability by allowing an attacker to compromise a less secure platform to gain passwords and user names and reuse on other systems. Users should be educated to choose unique passwords for critical systems, and systems should require users to change passwords periodically and disallow choosing historical passwords. Phishing attacks use social engineering, exploiting human psychology to gain access. These usually present as fraudulent email communications appearing to be sent from a trustworthy source asking the target to take some action, such as clicking on a link and entering sensitive information into a website made to look like a legitimate site or downloading malware. Some of the biggest cyberattacks in recent history like the Ukraine power grid attack in 2015 were launched initially with phishing emails \cite{ukraine}. Statistical evidence shows that nearly one-third of ransomware attacks happen due to a lack of cybersecurity training across organizations. If users are not aware of common cyber-threats and how they work, they will not be able to identify one and take precautionary actions at the right time. A single ransomware attack due to a weak user password or en employee accidentally downloading a malware software  can jeopardize the entire organization's security and its functioning. Therefore, organizations should make every effort to educate their employees on the basics of cybersecurity which includes spotting phishing emails or any unusual behavior of the system and how to choose and maintain credential information more wisely \cite{DATASHIELD_PROTECT}.

 Systems requiring lower assurance due to storing data at a lower sensitivity level may only require one authentication factor. Systems with higher assurance needs often require multi-factor authentication where two or more factors are combined to verify the identity of the user \cite{NIST_MFA}. To defeat multi-factor authentication, an adversary must compromise all authentication factors, requiring attacks against each authentication factor. For example, defeating a system that combines a credential card with a password and a voice recognition factor would require an adversary to gain physical proximity to the credential card in order to steal or copy it, obtain the password via password cracker or from a weaker system in the case of password reuse or plaintext storage or transmission of password, and obtain a high fidelity recording of the target user's vocal patterns. Clearly, defeating all authentication factors requires greater time and resources than defeating only one, changing the adversary model for this system versus a single-factor system. This also requires greater investment in constructing the security mechanisms by the defenders. Complex authentication mechanisms also increase the burden on users and overly complex mechanisms may lead to decreased adoption or compliance with users more likely to attempt to circumvent the mechanism. 

Authentication can also be performed as part of a cryptographic protocol wherein the source of a message is verified or as part of a key exchange protocol wherein the identity of the participants is verified. A message digest is a cryptographic hash of a message, called a digest as it represents the message in a fixed, usually smaller, size and is used as a primitive in protocols. When using a secure cryptographic hash function, the recipient independently computes $h(m)$ on the message $m$ received and compares to the hash to verify that the message has not been altered in transit. However, when both components are sent in the clear, an adversary with access to the network and ability to construct messages such as under the Dolev-Yao model can compute her own $m'$ and $h(m')$ which would pass the integrity check. A message authentication code (MAC) additionally requires the use of a shared secret such that only a participant in possession of the secret could compute the MAC. A number of algorithms can be used to compute MACs, such as symmetric key block ciphers, called CMAC, or keyed-hashing, called HMAC \cite{HMAC_NIST}. See \cite{MAC} for the three approved general purpose MAC algorithms as of June 2020.

\subsection{Authorization and Access Control} Authorization is ``the granting or denying of access rights to a user, program, or process'' \cite{NIST_assessment}. Generally, but not always, authorization follows subject authentication. Access control policies determine which subjects should be granted access to which resources possibly under which conditions. Access control mechanisms are implemented to enforce authorization decisions, protecting objects from unauthorized operations. There are four types of access control with some overlap utilized in information systems depending on the requirements of the system \cite{SP80053r5}. 

In mandatory access control (MAC), the policy determines which subjects may access which objects and subjects may not pass on granted access rights to unauthorized subjects. Additionally, subjects may not make changes to the policy, either by modifying access rights given to other subjects, existing objects, or new objects. The best known example is Department of Defense multi-level security where information is marked with a level and a subject with an appropriate clearance level may access the information. The Bell-LaPadula model was designed to formalize this system and has two properties for read and write access. The simple security property states that a subject is allowed read access if the subject’s level is higher than the object’s level, preventing a subject from accessing information he does not have the authorization to view. The star property states that a subject is allowed write access if the object’s level is higher than the subject’s, preventing leak of information of a higher security level from appearing in an object of a lower level. Conversely, discretionary access control (DAC) enables subjects to use their discretion with objects they have been granted access, such as by passing on their rights to other subjects or changing attributes of existing or new objects. DAC can be used with MAC; for example, the Bell-LaPadula model combines a DAC rule with the two MAC properties mentioned above. The access control mechanisms used to implement DAC policies are typically access control lists (ACLs) which create a mapping from subjects to permissions, typically on an identity basis, with one list per object while non-discretionary access policies generally use rule-based controls \cite{SP800192}. For example, in military security, security level rule matching is used to determine access, not a list of authorized subjects. An access control matrix (ACM) is an alternative to ACLs, stores the object permissions but on a per subject basis. 

Role-based access control (RBAC) arose from the ill-fit of the military implementation of MAC for civilian and commercial systems and weakness of DAC that enabled subjects to pass on rights to others \cite{ferraiolo1992role}. Additionally, ACL and ACM access control mechanisms are difficult to administer; accesses must be updated on a per subject and per object basis as subjects enter or leave the system and gain or lose privileges, creating opportunities for misconfiguration leading to vulnerabilities. RBAC assigns subjects to pre-defined roles that typically represent job functions within an organization and also assigns object permissions to roles. Thus, subjects acquire permissions based on the roles they hold. Access decisions consider whether the requested permission is allowed for some role held by the requesting subject. The role is thus an intermediary that simplifies management, and administration includes assigning or removing roles from subjects and giving or revoking permissions to roles by the object owner. A weakness of RBAC is that granularity of permissions is limited to the granularity of the role; all subjects assigned to the role inherit all permissions granted to the role, which may introduce security risks when the level of access is not necessary for the subject. Attribute-based access control (ABAC) assigns attributes to subjects and attributes to objects  \cite{SP800162}. Access control decisions are made by evaluation of a set of policies specified in terms of the attributes of the requesting subject, attributes of requested object, and possibly environmental conditions such as time of day or location. That is, the inputs to the decision engine are the attributes of the requesting user, attributes of requested object, environmental conditions, permission requested such as read or write, and rule set containing the policies. If a policy is found that is satisfied, access is granted. Without the administrative burden of constructing and maintaining ACLs per object or the potential mismatch of moving through intermediary roles, ABAC enables specification of fine-grained policies. 

Access control vulnerabilities arise in a number of ways. In MAC, enforcement stops at the boundary of the system; classified information leaks are examples of information failing to be protected outside of the boundaries. Weaknesses of DAC include that granting access is transitive and decided by the object owner rather than via system-wide policies and that access permissions inherit the identity of the subject executing the operation impacting auditability. Additionally, incorrect policy specification and conflicting rules can lead to denial of authorized subjects or leaking of information to unauthorized subjects. Methods for verifying and testing access control policies are discussed in \cite{SP800192}.

\subsection{Zero Trust Architectures} 
Previous security approaches to protect the data owned by an organization, either at rest or in transit, relied on perimeter defenses to keep attackers outside of the enterprise network; however, the complexity of modern enterprise environments leads to a perimeter that is often difficult or impossible to define. Perimeter based security suffers from lack of defense in depth as once an attacker breaches the perimeter or due to insider attack, no further mitigation is possible to prevent lateral movement across the network. Last, perimeter-based security cannot protect data residing outside of the network. Many modern use cases arise commonly where a different paradigm is needed such as an enterprise network with disconnected satellite networks or remote workers, cloud services hosting data and applications external to the network, visitors or contractors requiring limited access to enterprise resources, and 
cross-enterprise collaboration using a federated ID management system to allow authorized subjects from one enterprise to access resources on another.
Where the perimeter-based security approach establishes trust on the basis of network location, the zero trust paradigm requires trust to be established on a per-transaction basis \cite{rose2020zero}. Zero trust assumes the presence of an attacker already inside the network and as such does not distinguish between an inherently trusted internal environment and a non-trusted external environment. Additionally, it assumes that: devices on the network may not be under the enterprise's control due to bring-your-own-device (BYOD) scenarios from employees, contractors or guests; resources may be compromised; enterprise resources may not reside on enterprise-controlled infrastructure due to cloud hosting and remote work; and, local network connections such as those on which remote workers are operating from are untrusted. Due to the flow of data resources and assets across the ill-defined enterprise network perimeter, zero trust aims for a consistent security policy and posture across enterprise and non-enterprise infrastructure. To maintain data security in a compromised network, zero trust recommends that each request should have a preceding authorization decision based on  authentication and the resulting access should be based on need to access and with the least privileges granted to perform the duty. Zero trust architecture is the framework for using these principles combined with a risk-based approach to design an architecture given an enterprise's specific needs, assets, workflows, and risks. The following seven tenets of a zero trust architecture are provided in the NIST Special Publication 800-207 \cite{rose2020zero}:
\begin{enumerate}
    \item ``All data sources and computing services are considered resources.'' Thus, anything accessible on the network, including connected personal devices should be accounted for under the security strategy.
	\item ``All communication is secured regardless of network location.'' All communication requires confidentiality and integrity and should provide source authentication. These are typically achieved via encryption and cryptographic protocols.
	\item ``Access to individual enterprise resources is granted on a per-session basis.'' Trust must be established before each authorization decision, typically via authentication, and least privilege is used for access. Sessions can be defined as needed based on the specific resource and risk level. Granting access to one resource does not automatically grant access to another. 
	\item ``Access to resources is determined by dynamic policy—including the observable state of client identity, application/service, and the requesting asset—and may include other behavioral and environmental attributes.'' Policies are defined over subjects, resources, and actions. The requesting subject description can include information about the client identity, device used to request access, behavioral attributes such as patterns of use, and environmental attributes such as time, location, and current security threats on network.
	\item ``The enterprise monitors and measures the integrity and security posture of all owned and associated assets.'' In addition to evaluating trust at the time of request, continuous diagnostics and mitigation (CDM) can be deployed to monitor assets on the network, take action to patch or isolate those found to be compromised or vulnerable, and report issues.
	\item ``All resource authentication and authorization are dynamic and strictly enforced before access is allowed.'' Continuous monitoring to trigger reauthentication and reauthorization can be conducted based on policy inputs that consider aspects such as session length, suspicious or increased risk behaviors based on access requests or modification of resources but also balancing security, usability, availability, and cost.
    \item ``The enterprise collects as much information as possible about the current state of assets, network infrastructure and communications and uses it to improve its security posture.'' Instead of static rule-based policies, this information can aid in establishing suspicion of abnormal behavior for requests.
\end{enumerate}

Logical components \cite{rose2020zero} of a zero trust architecture are employed to achieve these tenets. A policy enforcement point (PEP), coming after a policy decision point (PDP), demarcates the boundary of the implicit zone trust and is the point after which no further action can be taken. The PEP is moved closer to a resource, shrinking the implicit trust zone. The PDP is composed of a policy engine (PE) that takes a policy and inputs from the environment to make and log access decisions using a trust algorithm as well as a policy administrator (PA) that executes the decision such as by configuring connections or issuing session tokens which it passes to the PEP. The inputs to the PDP can include: certificates from an enterprise PKI; requestor information such as identity records, roles, access attributes, and assigned assets from an ID management system; information about the state of the asset such as the patch status of the OS, any vulnerabilities, integrity of its components, and whether it has unauthorized components per a CDM program; attributes of the resources and data access policies; policy rules needed to remain compliant with an industry regulatory regime; network traffic and asset request logs; collected security information from a security information and event management (SIEM) system; and information about new vulnerabilities or active attacks from threat intelligence feeds.
The trust algorithm can vary from criteria-based which is generally static and composed by a human to a more dynamic one with computed confidence score and decision threshold. Additionally, it may be used to make a single access local decision that may fail to detect attacks or be done with a more global view of subject behavior given contextual knowledge of other decisions, requiring the PE to maintain a history as well as receive info from other PAs and PEPs across the network. 
The PEP which acts as a gate between the requestor and the resource on the data plane passes the request to the PDP on the control plane and then establishes, monitors, and terminates the connection from the PA's response.
Authentication coupled with environmental and behavioral attributes to determine if authorization is reasonable given confidence in the requestor's identity to achieve fine-grained access control designed for least privilege is performed at each PEP and therefore the temporal delay of performing these checks must be minimized for quality of service and balanced against risk.

Three approaches may be used in combination to implement a zero trust architecture \cite{rose2020zero}. The enhanced identity governance–driven approach determines resource access based on access privileges granted to given subject. It is often used with an open network model where network access is granted to all but resources are restricted by subject, especially when assets are not under the control of the network administrator such as with cloud storage. This open network access means that attackers can conduct reconnaissance and launch availability attacks, so other defenses must be in place to respond. The micro-segmentation approach places groups of similar resources on a small segment of the network with a gateway device such as firewall and uses PEP to dynamically allow access to the segment. It is in effect perimeter-based with a very small perimeter and does not allow fine-grained access on per resource basis to resources within the segment. Another approach uses a network overlay often with concepts from software-defined networking. After the access request is made to the PEP, the PA acts as a controller and reconfigures the network based on the PE decisions, and the PEP creates a secure channel between the requester and the resource. Deployment variations take into account differences such as: whether all devices will be owned by the enterprise and can have software installed vs. allowing non-enterprise owned devices (BYOD) that must obtain access through a portal; whether sandboxes can be installed on devices to host trusted applications that connect to the PEP; and whether gateways can be installed directly in front of a resource vs. in front of an enclave of resources such as in the case of legacy resources that cannot communicate directly with a gateway.

Network requirements to support a zero trust architecture can be summarized from  \cite{rose2020zero} as reachability, observability, and scalability. Reachability requirements are that assets can connect to the network and reach the PEP and resources should not be reachable without going through PEP, though the policy may hide some PEPs from some assets.   
Observability requirements are that the data and control planes are logically separate, the PEP is the only component that can access the PA, and the enterprise can observe all network traffic and knows which assets it owns and the devices' current security postures. Scalability requirements are that the remote enterprise assets can access resources without traversing the enterprise network and the infrastructure to support zero trust must be able to process the load placed on it given the additional activities for reauthorization and data collection.

Though an improvement over perimeter-based security, zero trust architectures are not free of threats \cite{rose2020zero}. As the PE and PA are essential to operation, they become a high value target and potential single point of failure for the resources they protect. Misconfiguration by an administrator or compromise via an attack leaves resources unprotected, which can be mitigated by careful configuration and monitoring. Availability attacks against the PE/PA leaves resources unavailable or a path disruption means a connection cannot be established, which can be mitigated by securing access to or replicating the PE/PA. Resources to which access has already been granted can be accessed via compromised credentials or accounts, which can be mitigated by reauthentication actions and MFA or a contextual trust algorithm that considers more than just authentication credentials. Securing communication across the network means some traffic may be encrypted especially if originating outside of the network, limiting observability and deep packet inspection, which can be mitigated by using other information or metadata about traffic flows along with ML to detect suspicious traffic. Logged contextual information and management policies can provide a wealth of information for an attacker to learn to avoid detection, which can be mitigated by protecting the log database and allowing access only to the most privileged admin accounts. Dynamic access decisions rely on data and without standards, the format may be vendor proprietary with a high cost of vendor switching, posing a problem in case of an attack that leaves a vendor compromised, which can be mitigated by holistic evaluation of vendors including issues like supply chain risk management and security. Last, the 
use of AI replacing human administrators in zero trust architecture needed for dynamic policy decisions based on continuous monitoring and global contextual information increases the risk of errors when the AI is not robust. This can be somewhat mitigated by utilizing human administrators to review and correct the decisions made by AI.

\subsection{Protecting Data at Rest}
Data at rest can be considered as the data that is not actively moving within multiple devices in a network or not being shared among devices or platforms in different networks. Properly implementing access control mechanisms and designing and verifying access control policies to correctly categorize data sensitivity and the necessary and sufficient subject access privileges as well as addressing software vulnerabilities that could be exploited by an attacker are some factors that need to be considered to ensure confidentiality of data at rest. NIST report \cite{black2016dramatically} defined a list of specific technical approaches that include improved methods of specifying, designing, and building software and better intrusion detection system to prevent or detect software vulnerabilities. 

Additionally, adopting modern encryption techniques like the Advanced Encryption Standard (AES) on hardware devices and in public cloud platforms used to store data protects data with a defense in depth strategy. When encryption keys are properly stored separately from the device or the data and strong encryption is used, accessing the plaintext information would require exceptional computational time and power even if the device is captured or the data management platform is compromised through software vulnerabilities or access control failures. Encryption algorithms are typically chosen so that the computation time to break the encryption exceeds the time for which the data needs to be protected. Quantum computers, once available, would have sufficient computational power to break most of today's encryption algorithms. In 2022, NIST recommended four post-quantum algorithms (CRYSTALS–KYBER, CRYSTALS–Dilithium, FALCON, and SPHINCS+) as the result of its six-year Post Quantum Encryption Standardization Process \cite{alagic2022status}; however, it is not currently necessary to abandon current algorithms like AES as it is expected to be secure for decades. Further, NIST does not recommend doubling the AES key length at this time, and provides guidance to use 128, 192, or 256 bit key lengths \cite{PQCFAQ}. Longer keys provide greater security but add processing time, so key length should be chosen based on the sensitivity of the data.

Cloud services are commonly used for data storage or processing when organizations require the flexibility of having access to these services based on changing data needs without the overhead cost of building and maintaining their own data centers. This creates a scenario where data is housed outside of the organization and the data owner cannot maintain control of the data through access control mechanisms. Data stored in the cloud could be accessed by the cloud service provider's administrators. Additionally, multi-tenancy is common in cloud services which means that an organization does not typically receive fully dedicated storage or processing resources and, as such, their data resides on the same servers as other customers. Misconfigurations of the platform's operating system or vulnerabilities within the cloud platform that allow memory allocated to one tenant to be accessed by another or failure to clear deallocated memory thus could lead to unauthorized data leakage. For both of these reasons, it is advised that customers be in control of encrypting their own data before moving it to the cloud service platform \cite{hu2020general}. 

As an example, Google Cloud performs encryption on customer data at rest as a default using primarily AES-256 \cite{googlerest} and  envelope encryption where data is encrypted by a data encryption key (DEK) and the DEK is encrypted by a key encryption key (KEK). In default encryption, Google and not the customer has access to the encryption keys. Data is broken into chunks and each chunk is encrypted with its own DEK and updates to a chunk of data are treated like a new chunk and encrypted with a new key. New KEKs are also created on a schedule with one current key for encryption and old keys available indefinitely for decryption. This key freshness is used to limit issues when a key is compromised. The chunks are distributed across storage so that an adversary will need access not only to all the chunks that correspond to the data but also to all the encryption keys that correspond to the chunks. DEKs are unique to the customer for resources not shared with other customers and non-unique for shared resources. Access control lists are used to limit the services that have access to the keys. In Google Cloud, the keys used to encrypt the DEKs are not unique to each customer; instead, they are shared across customers and unique to services. The argument is that having fewer KEKs than DEKs enables scalability, and management of keys is performed from their central keystore. Additionally, the KEKs do not leave the keystore, but  DEKs are decrypted within it and returned as plaintext to the storage or service requesting the key. Thus, a customer allowing protection of their data at rest to be handled by a cloud service provider who uses this model is trusting that the provider has properly configured their access control lists to ensure only the necessary services have access to DEKs, that misconfigurations will not allow their data to be decrypted by a service performing an action for another customer using the shared KEK, and that DEKs cannot be intercepted while being returned. Google Cloud also provides the ability for customers to manage their own keys \cite{googleCKMS}. They provide multiple granularities of customer control over encryption keys ranging from no customer access and management only by Google Cloud, customer access to keys generated and stored within the cloud, to keys generated and stored external to the cloud that are never sent to Google \cite{googleEKM}.

In addition to protecting confidentiality of data at rest, data integrity and availability must be protected. Access control is one mechanism for ensuring that unauthorized changes to the data cannot be made. However, this will not prevent accidental or intentional errors made by subjects with appropriate access permissions. Checksums and stored hashes may be used to catch errors due to corruption. Version control with backups and change logs can be used for maintaining the history of changes to data and conducting after the fact audits. Changes to highly sensitive data may warrant the use of digital signatures for tracking data authenticity. Information system resilience is "the ability of an information system to continue to: (i) operate under adverse conditions or stress, even if in a degraded or debilitated state, while maintaining essential operational capabilities; and (ii) recover to an effective operational posture in a time frame consistent with mission needs'' \cite{SP80039}. Maintaining access to critical data in the presence of cyber attacks such as denial of service or during natural events or infrastructure attacks that affect information systems such as power outages or even due to natural fluctuations in system use typically requires data storage and network path redundancy. Multiple paths through the network enable reachability of data when a segment of the network is overloaded or attacked. Redundant storage should reside on separate systems that would not be impacted by the same attacks or failures. At the same time, the redundant data must also be protected which adds both to the attack surface and to the complexity needed to manage the data, such as maintaining additional keys and access controls. For example, in Google Cloud, backups are typically encrypted with their own DEK. Thus, data should be stored with redundancy appropriate to the importance of the data. 

\subsection{Protecting Data in Transit} 
When data is transferred across a network, the primary security goals are confidentiality and integrity. Integrity mechanisms like message authentication codes allow for a check that the data arrives at the destination unaltered. Digital signatures combined with integrity checks are used to authenticate the source of the data. Encryption is typically performed to achieved confidentiality. While encrypted data can be broadcast or sent over a network without establishing a connection, one typically wants to know whether the data is received by the intended recipient. Therefore, the protection of data in transit often includes encrypting the data before transmission, authenticating the sender and receiver, and verifying that the received data was not modified. Encryption for data in transit often uses asymmetric key exchange techniques like Diffie-Hellman in order to establish a shared symmetric key or session key, often using the AES algorithm, that can be used for the data encryption, as asymmetric key encryption is slower and more computationally expensive than symmetric key. For transmission over the internet, Hypertext  Transfer Protocol Secure (HTTPS) is commonly used. HTTPS extended Hypertext Transfer Protocol (HTTP) previously with Secure Sockets Layers (SSL) but as SSL is now considered insecure, Transport Layer Security (TLS) is typically used. 

TLS provides confidentiality protection against eveasdropping, message integrity protection against man-in-the-middle attacks, authentication of one or both endpoints, and protection against replay of messages \cite{SP80052}. TLS consists of three subprotocols: handshake, change cipher spec, and alert protocols. During the handshake phase, participants negotiate parameters such as the cipher suite which specifies which encryption algorithms are used for different aspects of the protocol and to establish a shared secret used for deriving session keys by the endpoints such as the write key used for the client to send messages and for the server to send messages. A list of NIST recommended ciphers for TLS is available \cite{SP80052}. Integrity checks on messages is provided by the HMAC algorithm determined by the cipher suite in all TLS versions or authenticated encryption with associated data (AEAD) available in TLS 1.2 where the same write key is used for confidentiality and integrity. Authentication of the server is required, while authentication of the client is optional. Anti-replay prevention is provided through monotonically increasing sequence numbers contained within the integrity checked message. The client and server authenticate to each other either explicitly through the use of digital signatures using public-key certificates or the server can implicitly be authenticated by the client using the public key from the server's certificate while establishing the shared secret. As only the holder of the private key can decrypt that portion of the secret, both parties computing the same secret is proof of the server's identity. Security of TLS is dependent on endpoints having strong and secure private keys and on the PKI. As authentication is about establishing trust in the identity of the other endpoint, certificates provided by the endpoints should not be automatically trusted. Certificate authorities (CA) sign the certificates of endpoints to bind the name and public key of the endpoint as given in the certificate. CA certificates that provide the public key for the CA's signature are installed on systems and thus provide the root of trust for certificates received from other parties. Authentication in TLS can be violated if an attacker can obtain a certificate in the name of the target server signed by a CA trusted by the client \cite{SP80052}.

\subsection{Protecting Data in Computation}
Access control provides some protection on data in computation. A subject who has both read and write permissions can access data, perform a computation, and write the altered data back to memory. Access control logs enable auditing of changes to the data. In the typical case, the data may be encrypted in storage and then moved to a trusted system where it is decrypted, processed, and re-encrypted before finally being transmitted to storage again. Utilization of cloud services poses a new challenge where the system on which computation is occurring is outside of the organization and potentially untrusted. Computation cannot generally be performed on encrypted data as mathematical properties between ciphertext and plaintext are not maintained; i.e., the ciphertext obtained by adding two encrypted numbers will generally not decrypt to the plaintext result of adding the corresponding plaintext numbers. 

Homomorphic encryption algorithms are designed to allow mathematical operations to be performed on encrypted data without leaking any information about the corresponding plaintexts and without requiring the key or keys. Asymmetric encryption RSA does have the coincidental property of multiplicative homomorphism -- that the ciphertext of multiplied plaintext is the same as the multiplication of the ciphertexts of the plaintexts -- due to it's algebraic structure. However, RSA's homomorphism does not extend to computing over general functions, which was proposed but not solved by Rivest, Adleman, and Dertouzos in 1978 \cite{rivest1978data}. Gentry proposed the first fully homomorphic encryption (FHE) scheme that is homomorphic over arbitrary functions in 2009 \cite{gentry2009fully}. Homomorphic encryption schemes are categorized by the subset of the following attributes they possess \cite{armknecht2015guide}: 
\begin{itemize}
\item correct decryption where the scheme always produces the correct plaintext for a ciphertext; 
\item correct evaluation where the scheme correctly decrypts the ciphertext that is the result of the evaluated encrypted plaintexts to the same result that would be achieved by evaluating the plaintexts themselves with bounded probability for all functions provided for by the scheme;
\item compactness where the size of the ciphertext result is polynomial in the size of the security parameter; and
\item perfect, statistical, or computational circuit privacy where the distribution of ciphertexts resulting from evaluation over encrypted plaintexts and the distribution encrypting the resulting evaluations of plaintexts are indistinguishable either perfectly, statistically, or computationally, respectively.
\end{itemize}
The security parameter pertains to the key generation function. Somewhat homomorphic schemes possess correct decryption and correct evaluation and are not required to work over all functions, while FHE schemes possess correct decryption, correct evaluation, compactness, and works over all functions of arbitrary size \cite{armknecht2015guide}. 
There is a distinction between the distributions of so-called fresh ciphertexts or those resulting from encrypting a plaintext and evaluated ciphertexts or those resulting from evaluating a function on fresh ciphertexts, and the correct evaluation attribute is only required to hold over fresh ciphertexts. Performing a computation in some $i$ number of stages over intermediate ciphertexts, called $i$-hop homomorphic encryption, requires it to hold over ciphertexts resulting from evaluation up to $i$ stages as well. Theorems in \cite{armknecht2015guide} prove relationships involving perfect circuit privacy, staged computation, and scheme classification.

While the definition of FHE requires that computation over ciphertexts be efficient in the sense of being polynomial in the size of the security parameter, the polynomial is currently quite large for all FHE schemes and thus much slower on the ciphertext, making computing complex functions impractical \cite{armknecht2015guide}. Much of the current work to make FHE practical is designing faster algorithms and reducing the space requirements, and one of the few practical implementations is HElib \cite{halevi2018faster}. FHE techniques are typically based on adding noise to the ciphertext, and the noise stacks with each successive operation, restricting the number of computations that can be performed before decryption is no longer possible \cite{shruthi2021general}. Another challenge in utilizing cloud compute for processing while maintaining confidentiality is transitioning code that has been designed by the typical software developer to run on unencrypted data locally or within an organization's trusted environment to run on data encrypted by FHE in a cloud environment without extensive cryptography experience. Google's FHE transpiler bridges this gap, enabling conversion of code in a high-level programming language into a version of the code for FHE, with an open-source repository available for C++ \cite{shruthi2021general}.

\section{Data Privacy}
With the advancements in the ability to capture, store, and process large volumes of data, data-intensive applications are becoming more prevalent across domains. These applications analyze and identify underlying patterns in the dataset and assist in decision-making. However, throughout their lifecycle, data-intensive applications are susceptible to information leakages that could potentially expose sensitive private information about the individuals from the dataset. Thus, safeguarding the privacy of individuals from the dataset remains one of the primary challenges in data-intensive applications. From a privacy perspective, the key question is: how to extract useful and actionable insights from a dataset while protecting the privacy of individuals in the dataset?

Privacy-preserving data analysis is a set of approaches that are aimed at mitigating the risk of information leakage from a dataset that could potentially lead to accidental reconstruction and subsequent identification of individuals from the dataset. Anonymization techniques have been used for more than four decades to mitigate the risk of accidental disclosure of private information from a dataset after its public release \cite{domingo2021limits}. The primary goal of anonymization techniques is to prevent the re-identification of individuals in the dataset. A general approach to achieve this goal is to remove personally identifying attributes from the dataset prior to its release. However, studies from the literature demonstrated the feasibility of re-identifying individuals from an anonymized dataset \cite{ochoa2001reidentification,narayanan2008robust,rocher2019estimating}. 

Sweeney et al. demonstrated removing personal identifier attributes from a dataset does not necessarily guarantee user privacy \cite{sweeney2002k,samarati1998protecting}; specifically, it does not make the dataset anonymous. Given an anonymized health dataset and voting registry as auxiliary background information, they were able to re-identify individuals from the anonymized dataset using quasi-identifiers – a subset of attributes, when combined, can form an identifier that can lead to the re-identification of individuals from the dataset \cite{benschop2019statistical, carvalho2023survey, samarati1998protecting}. Similarly, Narayanan et al. successfully identified the users from an anonymized NETFLIX dataset by cross-referencing with a publicly available IMDB movie rating dataset \cite{narayanan2008robust}.

\subsection {k-anonymity}
To overcome the limitations of the traditional anonymization approaches, Sweeney et al. proposed k-anonymity, a privacy-preserving technique to prevent an adversary from re-identifying an individual from a dataset \cite{sweeney2002k,samarati1998protecting}. The k-anonymity technique involves a multi-step process for anonymizing a dataset. The k-anonymity technique begins with identifying quasi-identifiers from the dataset. As stated earlier, quasi-identifiers are a set of attributes from the dataset that can be used to identify an individual but are not unique to any individual. For example, in the case of a patient medical history dataset, the patient’s sex, age, and zip code are considered quasi-identifiers. These attributes can appear in other external datasets that are publicly available and can be easily leveraged by an adversary to re-identify an individual from the anonymized dataset. The second step in this process is to determine the value of k, which indicates the minimum group size for achieving k-anonymity. Next, in the third step, individuals or instances with similar characteristics or attributes are grouped together. If the total number of instances in a group is >= k, then the group is considered k-anonymous. However, if the number of instances in a group is less than k, it is not considered k-anonymous and is processed further, either by generalizing or suppressing the values of quasi-identifiers. The intuition is that a group with a sufficient number of instances that is greater than or equal to k makes it more challenging for an adversary to re-identify a specific individual or instance from the dataset. Once all the groups from the dataset satisfy the k-anonymity criterion, the dataset is considered a k-anonymized dataset. Overall, the k-anonymity technique guarantees that each individual or an instance will be indistinguishable from at least k-1 other individuals or instances in the dataset. Thus, making it harder for an adversary to re-identify an individual merely based on the identifiers. Furthermore, the value of k determines the level of privacy; a higher value of k increases the complexity for an adversary in re-identifying an individual or instance from the group. However, a higher value of k shall impact the utility of the data.

k-anonymity is a widely accepted technique for dataset privacy protection. However, in subsequent years, researchers discovered the limitations of k-anonymity techniques. Machanavajjhala et al. demonstrated that k-anonymity is vulnerable to background knowledge and homogeneity attacks \cite{machanavajjhala2007diversity}. They proposed l-diversity, an improvised privacy protection technique to address the limitations of k-anonymity, specifically attribute leakage. Later, Li et al. studied the limitations of l-diversity and proposed t-closeness, an approach to address the limitations of both k-anonymity and l-diversity \cite{li2006t}. While anonymization techniques such as k-anonymity, l-diversity, and t-closeness aim to deter privacy attacks, the fundamental idea of these approaches is to anonymize a dataset by grouping instances. This can result in information loss and, as a result, reduce the utility of the dataset. Additionally, the anonymized dataset might not be fully robust to background knowledge attacks.

\subsection{Differential Privacy}
The limitations of anonymization-based privacy-preserving techniques \cite{machanavajjhala2007diversity,li2006t} have emphasized the requirement for techniques that offer robust privacy protection guarantees. Dwork et al. proposed differential privacy, a formal mathematical framework to provide stronger privacy guarantees compared to k-anonymity and its family of approaches \cite{dwork2006calibrating}. The authors proposed the differential privacy framework to mitigate privacy risks in databases with the private information of individuals. 

Consider, for instance, a database consisting of health data records of "x" participants. As part of data analysis, an analyst queries the database for summary statistics such as the mean, maximum, and minimum for two attributes: height and weight. Let us assume a new participant is added to the study, and the revised summary statistics are published to the data analyst. In this scenario, by comparing the two sets of summary statistics, in addition to learning about the group, an adversary with sufficient background knowledge (awareness of the inclusion of a new participant) could potentially learn the new participant’s sensitive information (height and weight), thereby compromising individual privacy.

To mitigate such privacy risks, the differential privacy framework introduces privacy guarantees for interactive statistical queries by adding noise to the output of the database. In other words, when a practitioner or data analyst queries a database, the framework adds noise to the output, making it harder for an adversary to learn about an individual or instance from the dataset. Due to its superior performance compared to other privacy-preserving approaches, differential privacy has been widely adopted across various domains for use cases beyond its originally intended use case --- preserving privacy in interactive statistical queries. Notable examples include the use of differential privacy mechanisms by Apple \cite{appleDifferentialPrivacy,tang2017privacy}, Google \cite{erlingsson2014rappor}, and Microsoft \cite{ding2017collecting} for collecting user feedback. Another prominent application of differential privacy technique can be observed in the publication of US Census 2020 data to protect the privacy of individuals participating in the census \cite{ding2017collecting}.

Next, we present a brief overview of differential privacy. Given two neighboring datasets A and A’ differing by only one instance, a randomized mechanism M guarantees $\epsilon$-differential privacy, and, for every set of outcomes S, if M satisfies 
\begin{equation*}
Pr[M(A) \in S]\le exp(\epsilon)\times Pr[M(A') \in S]
\end{equation*}
$\epsilon$ is a measure of privacy loss, referred to as privacy parameter or privacy budget. A smaller $\epsilon$ value amounts to lesser privacy loss, thus indicating stronger privacy. Although stronger privacy is desirable from a privacy perspective, it could result in a loss in data utility. In other words, introducing noise to strengthen the privacy of a dataset could affect the usefulness of the data and potentially render the dataset utility less. In the context of data analysis, utility refers to learning or inferencing valuable insights from the data.  Practitioners aim to strike a balance between the privacy and utility tradeoff.

Traditional differential privacy mechanisms, in certain cases, could enforce stricter privacy restrictions by applying a large amount of noise to the data. Thus, inadvertently affecting the data’s utility and subsequently limiting the ability to learn or infer from the data.  As a result, due to its stricter privacy mechanisms, traditional differential privacy might not be applicable across different scenarios. To address this limitation, Dwork et al. proposed relaxed differential privacy \cite{dwork2006our, dwork2008differential}, and it is defined as follows: Given two neighboring datasets A and A’ differing by only one instance, a randomized mechanism M guarantees ($\epsilon, \delta$)-differential privacy, and, for every set of outcomes S, if M satisfies

\begin{equation*}
Pr[M(A) \in S]\le exp(\epsilon)\times Pr[M(A') \in S] + \delta
\end{equation*}

where $\epsilon$ represents the privacy budget, and  $\delta$ represents the strength of relaxation in differential privacy. In other words, $\delta$ represents the probability that an adversary can learn about the data. If the value of $\delta$ = 0, it results in a stronger notion of differential privacy. The relaxed differential privacy, ($\epsilon, \delta$)-differential privacy, provides a weaker privacy guarantee than the traditional $\epsilon$-differential privacy. However, the better practicality of the relaxed differential privacy with relatively less noise makes it a preferred privacy-preserving mechanism for machine learning applications. Discussion on the application of differential privacy in machine learning is presented in Section 4.3. 

\textbf{Sensitivity:} Recall that, for interactive statistical queries, the differential privacy framework aims to mask the differences between the responses from A and A’ by introducing a controlled amount of noise. The sensitivity of a query is a measure of the maximum possible change in the output when an instance is included or excluded from the dataset.  Based on the sensitivity of the query, the framework determines the maximum noise required to mask the outcome. The most commonly used method involves adding noise sampled from a Laplace distribution. Another option is to use noise that is sampled from a Gaussian distribution. However, both Laplace and Gaussian distributions might not be suitable for non-numeric data. In such scenarios, the exponential mechanism is used, which adds noise to the data by sampling from an exponential distribution. Regardless of the mechanism, the amount of noise added is proportional to the sensitivity of Mechanism M and is introduced in a way that ensures privacy.

\textbf{Composition:} The composition property of differential privacy suggests that executing n differentially private queries on a dataset weakens the overall privacy guarantees by a factor of n. For example, if three differential private queries with privacy budget $\epsilon$1, $\epsilon$2, and $\epsilon$3 are executed on the dataset, the resulting overall privacy budget would be $\epsilon$1 + $\epsilon$2+ $\epsilon$3.  Thus, executing three queries would provide a ($\epsilon$1 + $\epsilon$2+ $\epsilon$3)-differential privacy guarantee. In contrast to other privacy-preserving techniques, such as k-anonymity, which lacks the composition property, the differential privacy framework can defend against linkage attacks. In k-anonymity, adversaries can initiate multiple queries on a database, link their outcomes and potentially re-identify or de-anonymize an individual. On the contrary, in the differential privacy framework, when multiple queries are executed on a database, the noise injected by the differential privacy framework for each query is independent of the noise injected for another query. Thus, the framework guarantees privacy on executing multiple queries to a dataset. It is crucial to emphasize that executing multiple differentially private queries on a dataset can still weaken the overall privacy guarantees.


The literature encompasses a wide array of approaches aimed at safeguarding data privacy \cite{carvalho2023survey}. Within the scope of this manuscript, we present the two extensively used privacy-preserving approaches: k-anonymity and differential privacy. Given its characteristics, differential privacy is widely acknowledged as the preferred mechanism for privacy protection. Through the injection of noise, the differential privacy framework raises the complexity and difficulty for an adversary attempting to re-identify or de-anonymize individuals from a database. However, it is essential to recognize that, like any defense mechanism, differential privacy has its limitations. While differential privacy enhances privacy protection, its efficacy depends on a set of factors and striking a balance between privacy and preserving the utility of the data. An adversary with access to a large volume of data and background information could breach the privacy protections provided by differential privacy.

\section{Data Security and Machine Learning}

Data is one of the key components of ML. For a successful implementation and utilization of ML, it is essential to guarantee the safety and security of data used throughout the ML pipeline. Specifically, there is a need to establish robust measures that ensure data privacy protection across the ML lifecycle. This section presents an overview of the data privacy challenges across the ML pipeline and the current approaches to address those challenges.



\subsection {Data Privacy in Machine Learning}



The use of the term ``privacy'' varies between cybersecurity and ML literature. In cybersecurity, privacy pertains to the secrecy of individuals, focusing primarily on the confidentiality of personal information and safeguarding information from misuse or unauthorized access. In contrast, privacy in ML is more or less a synonymous with confidentiality, a term that is not typically used. As data can be considered to form the requirements of the ML-enabled system, data is foundational to the performance of an ML model. Thus, the primary focus of privacy in ML is protection of data with most emphasis placed on the training data. The direct application of differential privacy is clear when protecting the information of individuals whose data is used in the training process, but the same mechanisms seem to be employed when the sensitive training data does not pertain to individuals and would be considered confidential in the cybersecurity community. Test data also has a special need to be protected as one purpose of test data is to evaluate generalization to unseen instances. In competitions, whether academic or amongst competing vendors, the test data is withheld and protected to avoid leakage that compromises the utility of the test data. If the developers possess knowledge about the test data, it could lead to biasing the performance towards the test, either explicitly or implicitly. Thus, even when training data is made public, test data may be kept private.

ML-enabled systems are increasingly used across various domains, including safety-critical systems. These systems rely on ML models that are trained using large amounts of data, and their performance is directly attributed to the data they were trained with. In most cases, ML models are trained with sensitive private information belonging to individuals, which must be protected. To ensure the successful development and deployment of ML-enabled systems, privacy protection guarantees are necessary throughout the ML system's lifecycle. The primary objective is to ensure the protection of data, including sensitive information about individual data points from the training dataset, from any form of exposure or leakage throughout the ML lifecycle of the ML system. To address privacy concerns while harnessing the power of ML systems, privacy-preserving approaches have been proposed.

Privacy-preserving Machine Learning (PPML) is a set of techniques used to address privacy challenges in the ML lifecycle. ML research groups and industry practitioners have successfully adopted existing techniques from the fields of Statistics and Cryptography to enhance and guarantee data privacy across the ML lifecycle. Anonymization techniques such as k-anonymity, homomorphic encryption, differential privacy, and multi-party computation are some of the widely adapted techniques in PPML. Among these adapted techniques, differential privacy is the widely adopted approach in privacy-preserving machine learning \cite{jayaraman2019evaluating, domingo2021limits}.

\subsection{Privacy Challenges in ML Pipeline}
The lifecycle of an ML-enabled system can be broadly divided into three phases 1) the data collection and processing phase, 2) the model training phase, and 3) the post-model training phase. The first phase involves gathering, cleaning, sorting, and labeling the training data. In the second phase, the training data is provided as input to an ML algorithm, which analyzes and infers a decision logic based on the underlying patterns of the training dataset. This decision logic is referred to as an ML model and is further tested and validated. In the third phase, the validated ML model is integrated with the rest of the components of the ML system. This integrated ML-enabled system is then deployed and used for performing inference in the real world.

Malicious actors may attempt to extract sensitive information across the ML pipeline. Their objective is to either gather information about the dataset, including sensitive information, or information about the model, such as the model architecture, weights, parameters or both. For example, during the data processing phase, an attacker can attempt to access the dataset and learn whether certain individuals are part of the training data. In another scenario, attackers can frequently query a trained model and attempt to learn about the training dataset based on the model’s response to the queries. Rigaki et al. categorized privacy attacks in ML into four types: membership inference attacks, property inference attacks, model extraction attacks, and reconstruction attacks \cite{rigaki2020survey}.

\begin{itemize}

\item \emph{Membership Inference Attack: } Membership Inference Attacks (MIAs) aim to determine whether a data instance was part of an ML model's training dataset. In the post-training phase, an adversary may launch an MIA on pre-trained ML models and, if successful, could potentially obtain private sensitive information linked to individuals who were part of the training set. Based on the nature of the attack, MIAs can be either black-box or white-box attacks. Shokri et al. \cite{shokri2017membership} were the first to study membership inference attacks in ML. They demonstrated that in supervised learning models, the privacy of individual instances could be compromised through a black-box membership inference attack. Studies from the literature have showcased the efficacy of MIAs on supervised ML \cite{shokri2017membership}, federated learning \cite{melis2019exploiting, wang2019beyond}, and generative models \cite{hayes2017logan, hilprecht2019monte, chen2020gan}. Furthermore, \cite{long2018understanding} and \cite{yeom2018privacy} studied the relationship between overfitting and model susceptibility to MIA. Long et al. demonstrated that overfitting is a sufficient but not necessary condition for a successful MIA attack \cite{long2018understanding}. MIA is one of the most common types of privacy attacks in ML, and we refer the reader to \cite{hu2022membership} for a comprehensive review of different types of membership inference attacks on ML systems.

\item \emph{Property Inference Attack:} During the training process of an ML model, it unintentionally learns the statistical properties of the training data, including underlying distributions and aggregate information. Attackers can extract these features or properties of the training dataset by interacting with the ML model, referred to as a property inference attack. In certain scenarios, property inference attacks can leak sensitive information from the training data. For example, consider a facial recognition classifier trained predominantly with individuals belonging to a certain ethnicity. An attacker can use a property inference attack to learn this information, which was not intended to be publicized, and potentially skew the classifier's performance by exploiting this vulnerability. Recall that, a membership inference attack focuses on identifying whether a particular data instance was part of the training data or not. In contrast, a property inference attack aims to learn about the underlying attributes of the training dataset that the model learned unintentionally. Subsequently, an adversary can leverage the acquired information to exploit the model in a manner that compromises data privacy.

\item \emph{Model Extraction Attack: } A model extraction attack occurs in the post-training phase, where the attacker operates with no prior knowledge about the training data or the model, making it a black-box attack \cite{rigaki2020survey, liu2021machine, tramer2016stealing}. These attacks are carried out to either steal the functionality of a trained model and create a duplicate model that matches or exceeds the prediction performance of the original model, also known as accuracy extraction, or to steal information about the internal specifications of the ML model referred to as fidelity extraction, such as decision boundaries which can be exploited to launch reconnaissance-style attacks on the ML model in the future \cite{jagielski2020high}.

\item \emph{Reconstruction Attack or Model Inversion Attack:} Existing work from the literature \cite{dinur2003revealing, dwork2008new, dwork2017exposed} has demonstrated that it is possible to reconstruct or recover an individual’s data from aggregate statistical information. For example, the U.S. census bureau, in their dataset reconstruction experiments, out of 308,745,538 individuals from the 2010 census data, successfully re-identified 46\% of the individuals \cite{census2020, garfinkel2019understanding}. This example highlights that large-scale reconstruction attacks could inadvertently leak sensitive private information. ML-enabled systems are susceptible to reconstruction attacks because ML algorithms use statistical methods to infer underlying patterns from a training dataset. 

In their study, Fredrickson et al. in \cite{fredrikson2014privacy} proposed model inversion attacks. They demonstrated that given an ML model and partial knowledge of the training dataset, such as certain demographic information features associated with a patient, it is possible to predict the patient’s genetic markers, which are considered private information. Similarly, \cite{fredrikson2015model} exploits the confidence information of ML models and successfully reconstructs the training data. In \cite{balle2022reconstructing}, the authors demonstrated the capability of an adversary to successfully reconstruct the missing training data point by exploiting their access to the trained model and to all training instances, with the exception of one -- the missing training data point. Haim et al. proposed a method that reconstructs training data based on the parameters of a trained neural network classifier \cite{haim2022reconstructing}. They demonstrated that a significant amount of information about the training dataset is encoded in the model parameters, and this information can be leveraged to reconstruct the training samples. 

\end{itemize}

Given the data-intensive nature of ML systems, privacy attacks can occur across the ML pipeline. Sensitive private information of the individuals could be inadvertently revealed or leaked during the data collection or processing, during the training process, or in post-training model activities. Therefore, it is imperative to provide privacy guarantees across the ML lifecycle.  

\subsection{Differential Privacy for Machine Learning}

To address privacy challenges, differential privacy is utilized across the ML system lifecycle \cite{zhu2020more}. Practitioners employ differential privacy approaches at different stages of the ML lifecycle to establish privacy protections. Specifically, differential privacy is applied during the data collection and processing stage (Phase 1), model training phase (Phase 2), or at the time of model inference (Phase 3). Furthermore, the application of differential privacy at each phase provides varying levels of privacy guarantees within the ML pipeline.

\subsubsection{Phase 1: Data Collection and Processing}
In this phase, practitioners apply differential privacy to the training dataset prior to its use in the model training process. Typically, this is achieved by adding differentially private noise to the training dataset using one of two approaches listed below:
\begin{itemize}   \item\emph{Central Differential Privacy} – In a centralized privacy setting, data from all sources are collected and processed by a trusted aggregator, which is subsequently used in training the ML model. In this setup, the onus is on the aggregator to protect the privacy of the dataset. The aggregator employs differential privacy mechanisms to the curated dataset to ensure privacy. Differential privacy has been applied to anonymize heterogeneous data for classification tasks \cite{wang2020heterogeneous}, hierarchical datasets with sensitive information \cite{fioretto2021differential}, publishing or sharing sequential datasets \cite{chen2012differentially, chen2011differentially, mcsherry2010differentially}, vertically partitioned data \cite{tang2019differentially, mohammed2013secure}.

\item\emph{Local Differential Privacy} – As the name suggests, differential privacy is applied locally at the client-level to ensure privacy. In certain situations, there may arise a lack of trust among contributing users toward an aggregator's ability to safeguard privacy. Participants may also be uncomfortable sharing or processing sensitive information with the aggregator. In such scenarios, participants obfuscate their sensitive information by applying differential privacy before sharing it with the aggregator. Limited-precision local privacy (LPLP) \cite{schein2019locally} and privacy-preserving text perturbations \cite{feyisetan2020privacy} are some of the recent works that apply local differential privacy to guarantee data privacy.
\end{itemize}
Applying differential privacy at the data level is considered a complex task \cite{ponomareva2023dp} as it could inadvertently impact the utility of the model to perform its task, i.e., model performance. Adding excessive noise to the dataset can hinder the model's ability to learn, subsequently affecting the model's performance. Therefore, practitioners aim to strike a balance between privacy and model utility, aiming for a suitable tradeoff. In addition to its use in training datasets, differential privacy is applied in synthetic data generation to preserve privacy. By adding noise to datasets, differential privacy helps to guard against attribute inference attacks. For example, it can prevent an adversary from learning or inferring information about sensitive attributes from the dataset, making it an effective mechanism to deter attribute inference attacks.

\subsubsection{Phase 2: Model Training}
In this phase, a differentially private noise can be introduced at various stages. This can be done by injecting noise into the loss function, gradients, or model weights. Participants often share their model parameters with an aggregator in collaborative model learning. However, this sharing could lead to the inadvertent exposure of sensitive private information from the dataset on which the model was trained.  To protect the data privacy of each participant in the learning process and prevent unintentional leakage through the sharing of model parameters, Zhao et al. propose an approach that sanitizes model parameters by injecting noise into the objective function of the neural network\cite{zhao2019privacy}.

Another approach to guard against privacy leakage during the model training process is through output perturbation, where noise is injected into pre-trained model weights. This technique has been applied to trained logistic regression models \cite{chaudhuri2008privacy,zhang2012functional} and to a model trained in distributed learning environment \cite{jayaraman2019evaluating}. Injecting noise into the trained model weights helps mitigate both attribute and model inference attacks. However, one of the potential downside of this approach is the considerable impact on the overall performance of the ML model.

Among the various methods used to preserve privacy in the training process, gradient perturbation is the widely used approach for applying differential privacy in machine learning \cite{jayaraman2019evaluating}. The objective is to preserve privacy by introducing noise to clipped gradients \cite{abadi2016deep, shokri2015privacy} during training. Differentially Private Stochastic Gradient Descent (DP-SGD) is one of the widely used approaches to preserve privacy \cite{ponomareva2023dp}. Next, we present a brief overview of DP-SGD.

DP-SGD is an extension of Stochastic Gradient Descent (SGD), which is one of the popular approaches for training deep learning (DL) models. DP-SGD aims to preserve the privacy of the training data in a two-step process \cite{abadi2016deep}. 
First, using a technique referred to as norm clipping, if the L2 norm of any gradient exceeds a pre-defined threshold, it is clipped to the maximum norm. This technique prevents any single gradient from having a disproportionate influence on the model update, thus ensuring that the model is not too sensitive to a particular instance. In the second step, Gaussian noise is added to the gradients, which makes it harder for an adversary to infer sensitive information from the model updates. As in the case of any differential privacy mechanism, the amount of Gaussian noise depends on the privacy budget.  While DP-SGD is proven effective in preserving privacy, implementing DP-SGD can be computationally expensive, especially for large datasets. Furthermore, DP-SGD shall result in a significant drop in model performance. As \cite{ghazi2021deep} suggests, the accuracy of an ML model with DP-SGD (73\%) varies by more than 20\% compared to a non-private baseline (>95\%).

To overcome the computation cost of DP-SGD, Ghazi et al. proposed the idea of applying differential privacy to the labels of the instances used in training deep learning models \cite{ghazi2021deep}. In Label differential privacy (LabelDP), the assumption is that training instances are considered public while their respective labels are treated as sensitive private information, and their privacy is protected by adding differentially private noise. LabelDP is applied to regression \cite{ghazi2022regression} and deep learning \cite{ghazi2021deep}, and results from the literature suggest LabelDP is computationally easier to provide privacy guarantees than DP-SGD. Furthermore, LabelDP results in improved model performance compared to DP-SGD.

Among the three stages through which noise can be injected into the training process, injecting noise to the gradients is the widely adapted approach. Incorporating differentially private noise to the gradients or injecting noise to the loss function provides resistance against membership inference attacks but may not provide sufficient privacy guarantees against attribute inference attacks, as the noise is not introduced at the data level. Compared to these two training phases, introducing noise to the model weights offers better resistance against membership and attribute inference attacks. However, a potential drawback is the loss in the model’s utility. Since the model’s outcome is determined based on the weights, injecting noise into the weights could skew the model’s performance \cite{zhu2020more}. Overall, differential privacy help prevents privacy attacks during the model training process. In this phase, the noise is not directly injected into the data. As a result, in most scenarios, applying differential privacy during the model training phase does not significantly impact the model’s utility.

\subsubsection{Phase 3: Model Inference} In the post-training phase, differential privacy is applied to ensure that the ML model does not leak or expose private information during model inference. One approach to preserving privacy during the inference phase is to train the model first and then introduce noise to its outcome. Papernot et al. proposed private aggregation of teacher ensembles (PATE) for preserving privacy in the post-training phase \cite{papernot2016semi}. Given a sensitive dataset, PATE partitions the sensitive dataset into k disjoint training sets, where k is a specific number. Then, an ensemble of k teacher classifiers is trained using each disjoint training set. The ensemble classifier aggregates the predictions of the k teacher classifiers and adds a Laplacian noise to the outcome to preserve privacy. A student model is trained using a public dataset whose instances were labeled using the aggregate outcome from the teacher ensemble model. Finally, the student model, which was trained based on the knowledge aggregation of k ensemble classifiers, is used for model inference. Furthermore, PATE is extended to handle large-scale, imbalanced datasets \cite{papernot2018scalable}. The main objective of PATE is to decouple the sensitive information from the final model used for inference. The limitation of PATE is that it requires access to public data and, therefore, cannot be implemented in the absence of public data. Note that, differential privacy is applied during the inference phase of the ensemble teacher model, and the privacy guarantees are indirectly transferred to the student model used in real-world inference through the label generation process.

To summarize, the data-dependent nature of the ML-enabled system necessitates the need for strong privacy protections. Across the ML lifecycle, practitioners employ DP-based approaches to safeguard against privacy attacks by injecting noise into the data, loss function, gradients, model weights, labels, or model outputs. The amount of noise injected depends on the privacy budget, a user-defined measure of the desired level of privacy. In general, differential privacy is proven to be effective in guarding against various types of privacy attacks. However, implementing DP in the ML pipeline has its disadvantages in computational cost and privacy budget. Implementing DP-based approaches in the ML pipeline is computationally expensive, especially in the case of a large dataset. A larger amount of noise shall guarantee a better defense against privacy attacks, but it can inadvertently impact the utility of the model as excessive noise could hinder the model’s ability to learn effectively, resulting in suboptimal or poor model performance. Thus, choosing an optimal privacy budget is challenging and requires a tradeoff between privacy and model utility.

\section{Federated Learning}
One of the primary challenges in traditional ML is preserving the privacy of the data used in the model training activities. Given the nature of collecting, processing, and handling the data from a centralized location such as from cloud storage or a data server, the data is vulnerable to privacy attacks. To overcome this limitation, McMahan et al. proposed Federated Learning (FL), a cooperative approach to training ML and DL models \cite{mcmahan2017communication}. Federated learning  aims to alleviate the need to store and process data at a centralized location, for example, in a centralized data server. Instead, the data is captured, stored, and processed on the client side; the data is never transmitted to any external or other parties in the learning process.

Training a model in federated learning involves four phases \cite{wei2020federated, mammen2021federated}. First, a server or a central coordinator trains and broadcasts a global ML model to all the participating clients. Next, the ML model is re-trained over multiple iterations locally on the client side with the client's data. In the third step, each participating client sends their respective retrained model weights back to the server over a secured communication channel. Finally, the server aggregates the received weights from all participants, retunes the global model to perform model aggregation, and publishes the latest version of the global model to the participating clients.
The process is repeated until either of the two stopping conditions is satisfied: the accuracy of the global model exceeds the threshold, or the maximum number of iterations is satisfied. Note that the clients for the training process are selected randomly or using client selection algorithms [18]. Compared to traditional ML, a key aspect of federated learning is that all training data remains local on the client's device. In other words, the client is the sole owner of their respective data. As a result, the federated learning approach enables the development and deployment of personalized ML models while minimizing data privacy risks.

For example, consider the case of personalized next-word prediction in smartphones. One of the key challenges in building personalized ML models is handling a large volume of personalized user data while ensuring the protection of users' sensitive information. In such a scenario, federated learning can be leveraged as the data used for training the model is generated, captured, and processed locally and never transmitted or published outside the client's device. In this example, the aggregator, such as iOS or Android server, publishes a text predictor ML model to all the participating user's mobile devices, which are referred to as clients. Then, the ML model is fine-tuned based on the client's keyboard usage pattern. Next, the revised model parameters are sent back to the server; the server retrains the ML model using aggregated model weights and broadcasts the revised model to all the clients \cite{hard2018federated}. 

The decentralized data storage and distributed learning approach enable the training and deploying of ML models without the need for transmitting or exposing user data beyond their respective environments. Although federated learning, to a certain degree, is successful in thwarting data privacy attacks, the design and architecture of federated learning introduce a new set of challenges that results in the leakage of sensitive private information. In some cases, these attacks are targeted to extract or unauthorized access of private information of individuals used in model training activities \cite{bhowmick2018protection, zhu2019deep,fredrikson2015model,mcmahan2017learning,agarwal2018cpsgd}. Next, we discuss the privacy challenges associated with federated learning and explore the potential use of differential privacy as a means to address them. Note that while there exist additional challenges in federated learning, such as Scalability, Infrastructure, Resource Allocation, and Statistical Heterogeneity, this manuscript focuses solely on the privacy challenges of federated learning.

In federated learning, privacy attacks can originate from either the participants or the server in the learning framework. The collaborative learning mechanism of federated learning makes it vulnerable to malicious participants who may attempt to sabotage the overall learning process and, thereby, the performance of the ML model. Participants, either individually or as a group, may intentionally feed invalid or skewed model weights as updates to manipulate the models' performance. Additionally, the server itself may act in a compromised manner by exposing the private information of participating clients or jeopardizing the integrity of the learning process. Broadly, federated learning is susceptible to two types of attacks: poison and inference attacks \cite{lyu2020threats}. These attacks can occur either during the model training phase or the post-training phase.  

\begin{itemize}  \item\emph{Poison attack}: A participating client may modify or tweak their training data, resulting in skewed model performance. Moreover, since data is entirely owned and operated on the client side, there is a possibility that an adversarial participant could introduce or inject a stealth backdoor that could be exploited later to extract information, including the model information and possibly information about other participating clients in the training process.

\item\emph{Inference attack:} Recall that the main objective of a membership inference attack is to identify if a particular data record or an individual is part of the training dataset. Despite the use of decentralized data handling mechanisms, information about a participant or training instances can be inferred from the model updates shared between the participating clients and the server \cite{melis2019exploiting}. During training, the communication channel used to update model weights can be attacked to extract information about the training data and participants. Furthermore, Hitaj et al. demonstrated that a malicious participant in an federated learning setting can use a Generative Adversarial Network (GAN) to launch an inference attack, tricking other participants into releasing or sharing their sensitive private information \cite{hitaj2017deep}.
\end{itemize}

\subsection{Differential Privacy for Federated Learning}
Differential privacy has emerged as a promising approach to address the privacy challenges in federated learning. For example, in the case of mobile devices, differential privacy has been leveraged to safeguard user’s privacy in the next-word-prediction keyboards \cite{fl_keyboard}. Similarly, in the case of training Siri, a personalized AI voice assistant, user audio data has been anonymized using differential privacy to maintain privacy while enabling tuning the model for better performance \cite{Apple_Siri}.

One of the prevalent privacy concerns in federated learning is the possibility of exploiting the gradients of the model to extract information pertaining to the data instances or participants in the learning process. Geyer et al. applied differential privacy to protect the identification of participating clients in the federated learning training process \cite{geyer2017differentially}. This is in contrast to the application of differential privacy of machine learning, which primarily focuses on preventing the inadvertent disclosure or leakage of information pertaining to individual instances from a training dataset. Furthermore, Geyer et al. empirically demonstrated that with enough participating clients, the differential private federated learning model exhibits performance similar to that of the non-differentially private federated learning model. Agarwal et al. proposed a mechanism in which they added Gaussian noise to the model gradients before the client transmits its model parameters back to the server. By doing so, they aim to protect the identity of each participating client\cite{agarwal2018cpsgd}. Lu et al. proposed differentially private asynchronous federated learning (DP-AFL) for secure model updates by incorporating differentially private noise into the gradients during the training process on the client side\cite{lu2019differentially}. Truex et al. proposed LDP-Fed, a utility-aware privacy perturbation framework that applies differential privacy while minimizing utility loss  \cite{truex2020ldp}.  Differential privacy was applied to mitigate against reconstruction attacks \cite{bhowmick2018protection}. Liu et al. aim to address the communication overhead and privacy challenges in federated edge learning. They proposed an asynchronous differential privacy mechanism to implement node-level privacy and an asynchronous model update scheme to reduce communication overhead \cite{liu2021towards}.  McMahan et al. applied differential privacy to a federated averaging algorithm to guarantee user-level privacy. Given the scope of this manuscript, we limit our discussion to key differential privacy-based privacy-preserving mechanisms applied to federated learning \cite{mcmahan2017communication}. We refer the reader to \cite{yin2021comprehensive} for a comprehensive survey on different privacy-preserving mechanisms in federated learning.

In summary, while federated learning aims to address the traditional privacy challenges in ML via decentralized data handling and collaborative learning, participants, including clients or the server, may expose sensitive user information during the model training phase through weight update and gradient update operations, as well as during model inference \cite{orekondy2018gradient, qu2020decentralized, su2019securing}. To mitigate the aforementioned privacy challenges in federated learning, a widely adopted approach is differential privacy. Differential privacy adds a controlled amount of random noise to the training set or the gradients during training to safeguard against privacy attacks. This necessitates a complex trade-off between the utility of the model, specifically its accuracy, and the level of privacy. Additionally, differential privacy efficacy in safeguarding privacy is dictated by the size of the dataset. However, given the nature of federated learning, every participant is not guaranteed to train their respective model with a large dataset. This is because each participant trains their model with their respective dataset, which could vary in size. Consequently, this could affect the effectiveness of differential privacy \cite{enthoven2021overview}.

\section{Conclusion}
Though the field of cybersecurity has well established practices for protecting information, there is no one-size-fits-all approach. Data security must be designed for the specific needs of the information and the system in which it operates carefully considering aspects including sensitivity of the data, user awareness education and data use compliance, scalability trade-offs of performance or usability against security, use of cloud services for data storage or computation, and the structure of the organization's network, users, and resource distribution. In this survey, we have attempted to point the practitioner to basic information about these concerns, but each of them could easily be a survey of their own. 

In addition, the widespread and increasing use of ML presents new challenges for data security. ML is highly data dependent, causing an increase in the amount data that must be created, stored, processed, and protected for training these systems. Additionally, ML models are known to be brittle and often do not generalize well to new data, so high quality data matched to particular deployment use cases is extremely valuable, leading to a desire to both share data within and across organizations and yet also protect the data from unauthorized disclosure, damage, or destruction. Advancements in DP and FL attempt to address these challenges. This survey focused on aspects of data security assuming that the data exists in some information system, but additional weaknesses exist that can threaten ML-enabled systems requiring high quality data before the data is even collected. For example, an attacker that can manipulate a physical environment can threaten computer vision data collected for self-driving cars or an attacker can skew the distribution for presumably normal network traffic leading to classifiers that fail to properly detect her attacks. Outsourcing data labeling or using automatic labeling algorithms from open source repositories increase the attack surface. As ML-enabled systems are deployed with increasing frequency in critical systems, it is crucial to secure data across the entirety of the ML-operations pipeline and engage both cybersecurity experts and ML engineers in designing appropriate safeguards.

\begin{acks}
This material is based upon work supported, in whole or in part, by the U.S. Department of Defense through the Director Operational Test and Evaluation (DOT\&E) under
Contract HQ003419D0003. The Systems Engineering Research
Center (SERC) is a federally funded University Affiliated Research Center managed by Stevens Institute of Technology. Any views, opinions, findings and conclusions or recommendations expressed in this material are those of the author(s) and do not necessarily reflect the views of the United States Department of Defense nor DOT\&E.
\end{acks}


\bibliographystyle{IEEEtran}
\bibliography{references}


\end{document}